\documentclass[aps,prb,twocolumn,superscriptaddressgroupaddress,floatfix]{revtex4-2}
\usepackage{amssymb}
\usepackage{graphicx}
\usepackage{times}
\usepackage{amsmath}
\usepackage{amsthm}
\usepackage{amsfonts}
\usepackage[T1]{fontenc}
\usepackage[latin9]{inputenc}
\usepackage{array}
\usepackage{multirow}
\usepackage{color}
\usepackage{bm}
\usepackage{color}

\usepackage{bbm}
\usepackage{hyperref}
\usepackage{babel}
\usepackage{float}
\usepackage{hyperref}
\hypersetup
{	colorlinks,%
	citecolor=green,%
	linkcolor=red,%
	urlcolor=blue%
}
\allowdisplaybreaks[4]

\begin{document}
\title{Wiedemann-Franz law in graphene}
\author{Yi-Ting Tu}
\author{Sankar Das Sarma}
\affiliation{Condensed Matter Theory Center and Joint Quantum Institute, Department of Physics, University of Maryland, College Park, Maryland 20742, USA}

\begin{abstract}
  We analyze a well-known experimental work~[J. Crossno {\it et al.}, \href{https://doi.org/10.1126/science.aad0343}{Science {\bf 351}, 1058 (2016)}] which reported on the failure of the Wiedemann-Franz law in graphene at $T\sim 10\text{--}100\,\mathrm{K}$, attributing this failure to the non-Fermi liquid nature of the Dirac fluid associated with undoped intrinsic graphene.
 In spite of serious theoretical efforts, the reported observations remain unexplained.
 Our detailed quantitative analysis based on Fermi liquid considerations, which apply to extrinsic doped graphene, establishes that one possible explanation for the reported observations is the opening of a gap at the Dirac point, induced perhaps by the boron nitride substrate.
 We suggest that more experiments are necessary to resolve the issue, and we believe that the experiment may not actually have anything to do with Dirac fluid hydrodynamics, but relates to finite-temperature low-density bipolar diffusive transport by electrons and holes in the presence of short- and long-range disorder, and phonons.
\end{abstract}

\maketitle

\section{Introduction}

Crossno \emph{et~al.}\ published a high-profile experimental paper in 2016~\cite{Crossno2016}, entitled ``Observation of Dirac fluid and the breakdown of the Wiedemann-Franz law in graphene'', where they reported a measurement of the Lorenz number $L=\kappa/(\sigma T)$ in graphene as a function of doping density ($n$) and temperature ($T$), with $\kappa$ and $\sigma$ being respectively the carrier (both electrons and holes, but not phonons) thermal and electrical conductivity.
 It has been known for a long time~\cite{Franz1853} that ordinary metals at room temperatures obey the universal Wiedemann-Franz (WF) law where $L = L_0$, essentially a constant independent of the material and the temperature (as long as it is not very low), where $L_0$ is the so-called Lorenz constant, $L_0=\frac{\pi^2}{3}\left(\frac{k_B}{e}\right)^2$, with $k_B$, $e$ being the Boltzmann constant and electron charge respectively~\cite{Lorenz1881}.
 The WF law is widely obeyed by metallic Fermi liquid systems, where both energy and charge transport are carried by the same free carriers as long as inelastic scattering effects are unimportant~\cite{Mahajan2013,Lavasani2019,Ahn2022}.
 In fact, many strongly correlated systems (e.g.\ cuprate high-temperature superconductors), which are sometimes considered to be non-Fermi liquids, also appear to obey the WF law.
Since inelastic scattering typically vanishes at zero temperature, the WF law in all likelihood applies to all electronic systems at zero temperature.

The remarkable aspect of Ref.~\cite{Crossno2016} is that it reported a dramatic violation of the WF law, where the measured Lorenz ratio $L/L_0$ shows a large peak at the graphene charge neutrality point (CNP) which is nonmonotonic in temperature, manifesting a value as large as $\sim 20$ at $T\sim 60\,\mathrm{K}$, decreasing for  temperatures below and above.
 It is noteworthy that the experiment found $L\gg L_0$ in some region of $(n,T)$ since typically electron-electron scattering predicts a suppression of $L<L_0$ and not an enhancement~\cite{Ahn2022,Lucas2018}.
The totality of the various experimental features in Ref.~\cite{Crossno2016} has led to the somewhat ill-defined claim that the observations are consistent with the hydrodynamic quantum critical behavior of non-Fermi-liquid Dirac fluids, but no explicit calculations verify such strong claims.
It is well-established experimentally and theoretically that the graphene Dirac point is an unstable fixed point that is strongly suppressed by electron-hole puddles arising from Coulomb disorder in the environment~\cite{Adam2007}.
In particular, the predicted conductance quantization of the intrinsic Dirac point has never been observed in bulk graphene at the CNP, showing that the physics of graphene is associated with extrinsic doped graphene with the critical Dirac point being inaccessible experimentally because of charge puddles.
A detailed phenomenological hydrodynamical theory~\cite{Lucas2016} based on the Dirac fluid idea could not explain the data of Ref.~\cite{Crossno2016} despite having 6 free parameters in the theory~\cite{LucasPrivate}.
In particular, the puzzling peak in the Lorenz ratio for $T\sim 60\,\mathrm{K}$ reported in Ref.~\cite{Crossno2016} remains unexplained.

Given this unsatisfactory situation surrounding a high-profile publication~\cite{Crossno2016} with an important claim of the observation of the quantum Dirac fluid, we revisit the experiment using a more pedestrian approach assuming that the observed WF law breakdown in the experiment arises from the bipolar diffusion associated with the motion of both electrons and holes in the system at finite temperatures.
 We find that we can explain the observations of Ref.~\cite{Crossno2016} qualitatively and semiquantitatively within a bipolar diffusion Boltzmann transport model including disorder and phonon scattering effects if we assume that an energy gap has opened up at the Dirac point due to the effect of the hexagonal boron nitride (hBN) substrate, which is possible in principle~\cite{DeanPrivate,Hunt2013,RibeiroPalau2018,Finney2019,Yankowitz2018}.
Our work includes only short-range and long-range disorder, and acoustic phonons in the theory along with the full effects of bipolar diffusion by both electrons and holes in a gapped system.
 The goal is to see the extent to which the interesting WF law breakdown data reported in Ref.~\cite{Crossno2016} can be captured in a theory that assumes the system to be extrinsic graphene (i.e.\ doped graphene) with the carriers scattering from disorder and phonons as in simple metals.
 This is necessary given the failure of the interacting Dirac liquid theory in explaining the observations of Ref.~\cite{Crossno2016} even using multiple independent fitting parameters.

\section{Background}

Electron-electron interaction effects in graphene are well-understood and were calculated in depth in several earlier references~\cite{DasSarma2007,Li2013,Hwang2008,Barnes2014}, and we first discuss its relevance to the WF experiment of Ref.~\cite{Crossno2016}.
 As was pointed out very early~\cite{DasSarma2007}, there are two qualitatively different conceptual situations to consider: intrinsic graphene with no doping (i.e.\ just a hypothetical pristine graphene Dirac fluid) and extrinsic doped graphene (i.e.\ graphene with free carriers in the conduction or valence band depending on whether the doping is electron-like or hole-like).
Due to the invariable presence of random charged impurities in the environment, the graphene layer is dominated by electron and hole puddles around the CNP, and is effectively always spatially inhomogeneously doped. The Dirac point (and therefore, the Dirac fluid) is, therefore, inaccessible with the conductivity developing a plateau around the CNP with an approximate value of $5\text{--}50 e^2/h$, which is strongly dependent on the sample disorder and is larger than the predicted universal Dirac point quantum conductivity $\sim 4e^2/(\pi h)$~\cite{Adam2007,DasSarma2011,Adam2009}.
 The approximate region of the graphene minimum conductivity plateau in carrier density around the CNP depends on the sample disorder in a complex manner (because the puddle properties depend on the disorder details), and is smaller in general for cleaner samples.
In Ref.~\cite{Crossno2016}, the measured minimum conductivity is $\sim 8\text{--}12 e^2/h$ and the plateau size is $\sim 10^{10}\,\mathrm{cm}^{-2}$, which implies low sample disorder (which is consistent of the high-mobility and high-quality nature of the graphene on hBN samples used in Ref.~\cite{Crossno2016}).
The $\mathcal{O}(10^{10}\,\mathrm{cm}^{-2})$ puddle regime of Ref.~\cite{Crossno2016} implies an intrinsic Fermi energy of $>100\,\mathrm{K}$, which means that to access the Dirac point one needs $T\gg 100\,\mathrm{K}$, where phonon effects would become crucial.
 Thus, the intrinsic graphene (or equivalently, Dirac fluid) properties of the Dirac point are not accessible.
 We note that Ref.~\cite{Crossno2016} does not report the observation of the graphene universal Dirac point quantum conductivity, but nevertheless interprets the data based on the assumption~\cite{Kashuba2008,Fritz2008} of an intrinsic Dirac fluid hydrodynamics. 
 We believe that the absence of the universal Dirac point conductivity in Ref.~\cite{Crossno2016} implies that the system is dominated by disorder puddles and is not in the interaction-dominated hydrodynamic Dirac fluid regime.

\begin{figure}
  \includegraphics[trim=0 0 0 0, clip]{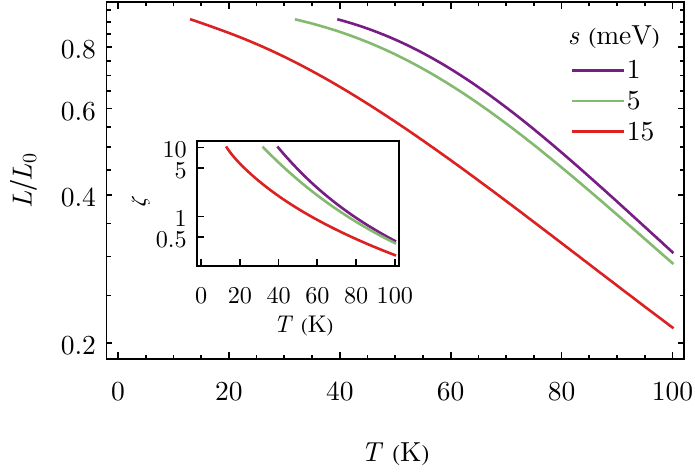}
  \caption{Shows the calculated dimensionless Poiseuille parameter $\zeta=\tau_\text{ee}/\tau_\text{imp}$ (inset) as well as the corresponding Lorenz ratio (main figure) for different puddle parameter values ($s$) at a fixed carrier density of $10^{10}\,\mathrm{cm}^{-2}$ as a function of temperature for extrinsic graphene following Refs.~\cite{Li2013,Ahn2022}.
The point to note is the monotonic suppression of $L/L_0$ with increasing $T$ in the regime where phonons must be accounted for.
}
  \label{fig:ee}
\end{figure}

 To investigate further the possible role of electron-electron interactions, we use Ref.~\cite{Li2013} to calculate the electron-electron scattering rate in graphene, comparing it with the electron-impurity scattering rate and obtaining the Lorenz ratio assuming extrinsic graphene (i.e.\ just electrons) by following \cite{Ahn2022}.
The results for different puddle disorders are shown in Fig.~\ref{fig:ee}.
 The most important aspect of this analysis for our purpose is that the parameter $\zeta=\tau_\text{ee}/\tau_\text{imp}$ which measures the ratio of the momentum-conserving electron-electron scattering to the momentum-nonconserving electron-impurity scattering exceeds unity only at a relatively high temperature ($T>60\,\mathrm{K}$) even in very clean graphene, thus implying that any graphene hydrodynamic effects would not show up for low $T$.
 Note that phonons are not included in Fig.~\ref{fig:ee} results.
 At such high temperatures, electron-phonon interactions cannot be neglected, and thus it is unlikely that pristine Dirac fluid effects of intrinsic graphene (where electron-electron interactions dominate over electron-impurity and electron-phonon interactions) can decisively manifest in monolayer graphene (just as this does not happen in simple normal metals).
 This becomes clear once we consider that the effective $T_\text{ph}$ (which is roughly $T_\text{BG}/5$ with $T_\text{BG} \sim n^{1/2}$ being the Bloch-Gr\"uneisen temperature), where phonon scattering becomes significant, is of $\mathcal{O}(10\,\mathrm{K})$ at low carrier densities of relevance in Ref.~\cite{Crossno2016}~\cite{DasSarma2011}.
 Basically, there is no effective window in the temperature-density-disorder-phonon parameter space where the Dirac-point-tuned intrinsic graphene non-Fermi liquid physics can dominate in controlling the Lorenz ratio of monolayer graphene~\cite{Ho2018}.

 One way of seeing this is that the inequality $T_\text{ph}\gg T\gg T_F\gg T_\text{puddle}$ (the energy scale of puddles), which is necessary for intrinsic graphene, is not easy to satisfy, due to the fact that $c_s$, the speed of sound in graphene, is much smaller than $v_F$.
 It is therefore incorrect to assume that a particular graphene sample is in the hydrodynamical Dirac fluid regime \emph{a priori} since the necessary conditions are highly restrictive both on the low- and high-temperature sides.
 This may be the reason why the theory of Ref.~\cite{Lucas2016} is unable to explain the Crossno experiment in spite of having multiple adjustable parameters.
 In addition, interaction effects tend to suppress (as in Fig.~\ref{fig:ee} where $L/L_0<1$ for higher $T$ where interaction dominates) the Lorenz ratio (i.e.\ $L<L_0$) in single-component extrinsic systems~\cite{Ahn2022,Lucas2018}, whereas the Crossno experiment reports a large enhancement ($L/L_0\sim 20$) at higher $T\sim 60\,\mathrm{K}$.
 This is unexpected and is an additional mystery.
 In fact, measurements on extrinsic doped graphene show good agreement between the experimental data and the WF law, which is expected based on the dominance of impurity and phonon scattering in graphene essentially at all temperatures~\cite{Kim2016}.

Given this background, we develop a Boltzmann-equation-based kinetic theory for the WF law in graphene, including short-range disorder, long-range disorder, and phonon.
It is well-known that long (short) range disorder dominates graphene transport properties at low (high) carrier densities, and phonons become relevant at higher temperatures~\cite{DasSarma2011,Adam2007,Chen2008,Tan2007}.
The crucial feature of the theory is that we include the bipolar diffusion effect quantitatively and nonperturbatively, treating both electrons and holes on an equal footing, leading to the enhancement of the Lorenz number as observed experimentally.
Our theory can account for most features of the observations (except of the very large peak in $L/L_0\sim 20$) of Crossno \emph{et al.}~\cite{Crossno2016}, and therefore, we generalize the theory to the existence of a gap opening up at the Dirac point by virtue of the hBN substrate~\cite{DeanPrivate,Hunt2013,RibeiroPalau2018,Finney2019,Yankowitz2018}.
 It is well-known that hBN could induce a gap in graphene, and the size of the energy gap depends on the details of how the layers are oriented and other configurational details.
 Using this gap as an additional parameter, we are able to qualitatively explain the Crossno observation of a peak in $L/L_0$ at $T\sim 60\,\mathrm{K}$.
 Of course, whether the sample used in Ref.~\cite{Crossno2016} has an energy gap is unknown and unknowable, and our inclusion of an energy gap in the theory should be considered as a model assumption which is certainly a possibility, but by no means a certainty.
Future experiments should check our predictions by ensuring a gap at the Dirac point.

In the next two sections, we present our theory  (Sec.~\ref{sec:theory}) and results (Sec.~\ref{sec:results}), and we conclude in Section~\ref{sec:conclusion} with a summary of our results and a discussion of open questions.
 Appendices \ref{sec:parabolic}--\ref{sec:singlelinear} provide additional results for different band structures for comparison, and Appendix \ref{sec:microscopic} provides the microscopic calculation results for completeness.
 We mention that by the very nature of our theoretical work, we provide extensive calculated results for the Lorenz ratio $L/L_0$ as a function of temperature, carrier density, energy gap, and scattering mechanisms with the totality of our results providing a context for understanding the intriguing experimental data presented in Ref.~\cite{Crossno2016}.

\section{Theory}\label{sec:theory}

We first divide our system into the conduction band (labeled with $+$) and the valance band (labeled with $-$), then calculate the electrical and thermal conductivity separately using the Boltzmann transport theory, and then combine them with the bipolar diffusion effect included. The value of $L/L_0$ is then obtained as a function of density and temperature for various scattering mechanisms.

In the ideal situation, the system is gapless, with the energy dispersion $\varepsilon_{\pm}(\mathbf{p})=\pm v_F |\mathbf{p}|$, where $v_F\sim 1\times10^6\,\mathrm{m/s}$ is the Fermi velocity of graphene~\cite{DasSarma2011}.
But we also consider the situation in which a gap opens up at the Dirac point. Since the exact dispersion near the gap is unknown, we consider the simplest model:
\begin{align}\label{eq:dispersion}
  \varepsilon_{+}(\mathbf{p})&=+ v_F |\mathbf{p}|\nonumber\\
  \varepsilon_{-}(\mathbf{p})&=-v_F |\mathbf{p}|-\Delta\,,
\end{align}
where $\Delta$ is the size of the gap. We show in Appendix~\ref{sec:parabolic} that parabolic dispersion near the gap shows similar results, establishing the universality of the simple model.
We include both spin and valley degeneracies~\cite{DasSarma2011} for the density of states:
\begin{align}
  D_+(\varepsilon)&=\frac{2\varepsilon}{\pi\hbar^2 v_F^2}\quad\text{for }\varepsilon>0\nonumber\\
  D_-(\varepsilon)&=\frac{2(-\Delta-\varepsilon)}{\pi\hbar^2 v_F^2}\quad\text{for }\varepsilon<-\Delta\,.
\end{align}

When there is an applied electrochemical force ($\mathbf{\mathcal{E}}$) and temperature gradient ($\mathbf{\nabla} T$) in the linear response regime, the electrical and thermal current can be written as 
\begin{equation} 
  \renewcommand*{\arraystretch}{1.5}
  \begin{pmatrix}
    \mathbf{J}_e^\pm\\
    \mathbf{J}_q^\pm
  \end{pmatrix}
  =
  \begin{pmatrix}
    L_{EE}^\pm & L_{ET}^\pm \\
    L_{TE}^\pm & L_{TT}^\pm
  \end{pmatrix}
  \begin{pmatrix}
    \mathcal{E}\\
    \mathbf{\nabla} T
  \end{pmatrix}\,.
\end{equation}
We use the extensively used Boltzmann-equation-based approach with the relaxation time approximation formulated in Ref.~\cite{Lavasani2019}, in which the transport coefficients are given by the formulas:
\begin{align}\label{eq:transport}
  L_{EE}^\pm&=e^2\int d\varepsilon\left(-\frac{\partial f_0}{\partial\varepsilon}\right)D_\pm(\varepsilon) v_\pm^2(\varepsilon) \tau_{\sigma}^\pm(\varepsilon)\nonumber\\
  L_{ET}^\pm&=\frac{e}{T}\int d\varepsilon\left(-\frac{\partial f_0}{\partial\varepsilon}\right)(\varepsilon-\mu)D_\pm(\varepsilon)v_\pm^2(\varepsilon)\tau_{\kappa}^\pm(\varepsilon)\nonumber\\
  L_{TE}^\pm&=-e\int d\varepsilon\left(-\frac{\partial f_0}{\partial\varepsilon}\right)(\varepsilon-\mu)D_\pm(\varepsilon)v_\pm^2(\varepsilon)\tau_{\sigma}^\pm(\varepsilon)\nonumber\\
  L_{TT}^\pm&=-\frac{1}{T}\int d\varepsilon\left(-\frac{\partial f_0}{\partial\varepsilon}\right)(\varepsilon-\mu)^2D_\pm(\varepsilon)v_\pm^2(\varepsilon)\tau_{\kappa}^\pm(\varepsilon)\,.
\end{align}
Here $e$ is the electron charge, $T$ is the temperature, $f_0=\left(1+\exp\left(\frac{\varepsilon-\mu}{k_BT}\right)\right)^{-1}$ is the equilibrium Fermi distribution function, $v_\pm(\varepsilon)=\partial\varepsilon_\pm/\partial p$ ($=v_F$ for our model here), $\tau_{\sigma,\kappa}^\pm$ are the electrical and thermal relaxation time, and the integrations are over the entire energy region of the corresponding band.
The chemical potential $\mu$ is obtained self-consistently from the carrier density
\begin{equation}\label{eq:consistent}
  n=\int_0^{\infty}d\varepsilon D_+(\varepsilon)f_0(\varepsilon)-\int_{-\infty}^{-\Delta}d\varepsilon D_-(\varepsilon)(1-f_0(\varepsilon))\,.
\end{equation}
In order to describe our system with dimensionless parameters, sometimes the Fermi temperature $T_F$ is used instead of $n$:
\begin{equation}
  n=\int_0^{k_BT_F} d\varepsilon D_+(\varepsilon),\quad\text{assuming }n>0\,.
\end{equation}
For graphene, the relationship between these two parameters is $T_F= (\hbar v_F/k_B) (\pi n)^{1/2} \sim 135 (n/(10^{10}\,\mathrm{cm}^{-2}))^{1/2}\;\mathrm{K}$.

The relaxation times depend on the details of the scattering mechanisms, and $\tau_\sigma^\pm$ (the electrical relaxation time) may or may not equals $\tau_\kappa^\pm$ (the thermal relaxation time).
By particle-hole symmetry, we can relate the relaxation times for the two bands as $\tau_{\sigma,\kappa}^+(\varepsilon)=\tau_{\sigma,\kappa}^-(-\Delta-\varepsilon)\equiv\tau_{\sigma,\kappa}(\varepsilon)$ for $\varepsilon\geq 0$.
In addition, we will mainly consider $\tau_{\sigma}(\varepsilon)=\tau_\kappa(\varepsilon)\equiv\tau(\varepsilon)$.
Typically, for a single scattering mechanism, we can approximate $\tau(\varepsilon)\sim\tau_0\varepsilon^j$~\cite{Hwang2009a}, where $j$ is the scattering exponent for that mechanism, and $\tau_0$ may depend on $T$ as well as $n$ (but not on $\varepsilon$).
When several mechanisms are combined, the total scattering rate $1/\tau$ equals the sum of the individual ones.
We will consider both pure as well as combined scattering mechanisms in this paper.
In Appendix~\ref{sec:microscopic}, we derive $\tau_{\sigma,\kappa}$ directly from the Boltzmann transport theory, explicitly verifying that this scattering exponent model is a reasonable qualitative approximation for all the microscopic scattering mechanisms of our interest.

After obtaining the transport coefficients, the electrical conductivity $\sigma_\pm=L_{EE}^\pm$, thermal conductivity $\kappa_\pm=-L_{TT}^\pm+L_{TE}^\pm L_{ET}^\pm/L_{EE}^\pm$, and thermopower $S_\pm=L_{ET}^\pm/L_{EE}^\pm$ can be calulated.
The conductivities of the total system are
\begin{align}
\sigma&=\sigma_++\sigma_-\nonumber\\
\kappa&=\kappa_++\kappa_-+\frac{\sigma_+\sigma_-}{\sigma_++\sigma_-}T(S_+-S_-)^2\,.
\end{align}
The third term of $\kappa$ is the bipolar diffusion effect~\cite{Goldsmid2010}, which is crucial in our theory. For comparison, we also do some of the calculations without bipolar diffusion by dropping this term. Finally, we obtain $L=\kappa/(\sigma T)$.
Note that the presence of bipolar diffusion, i.e.\ both electrons/holes participating in transport, enhances $L/L_0$ above unity in principle.

For notational brevity, we will directly use the Kelvin for the unit of energy (i.e.\ setting $k_B=1$) in the following sections.

\section{Results}\label{sec:results}

\begin{figure}
  \includegraphics[trim=5 15 0 15, clip]{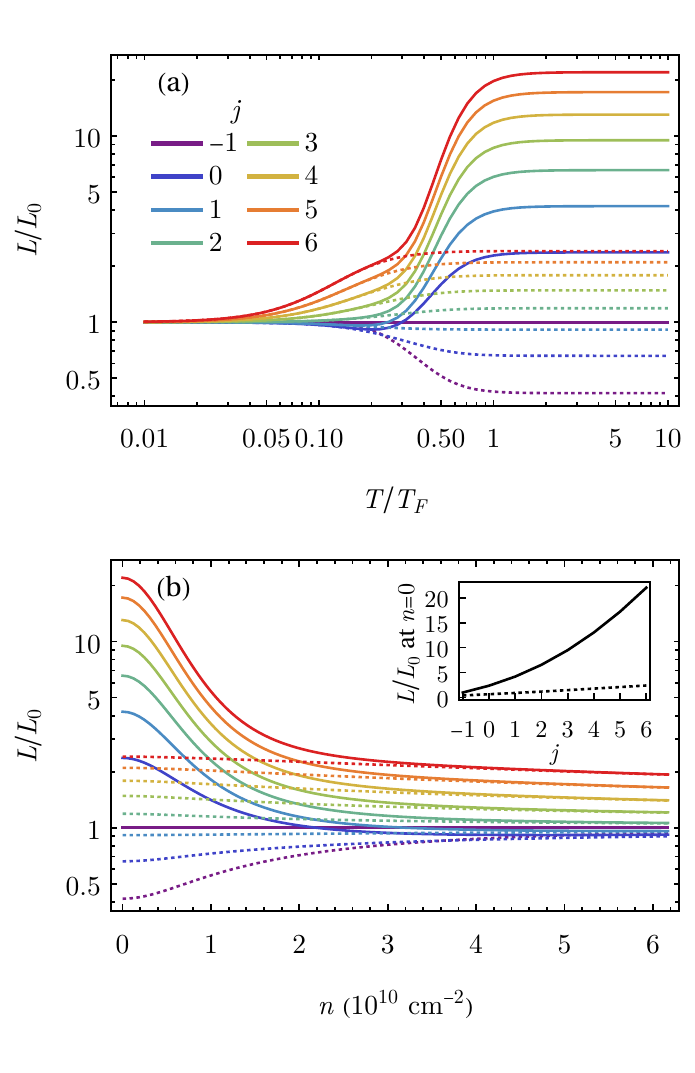}
  \caption{
    Shows the calculated $L/L_0$ for (a) fixed $n$ as a function of $T$ and for (b) fixed $T=60\,\mathrm{K}$ as a function of $n$ in the gapless linear dispersion model for a scattering rate $\tau(\varepsilon)=\tau_0\varepsilon^j$.
    The inset shows the value of $L/L_0$ at $n=0$, which equals the saturated high-$T$ nondegenerate ($T\gg T_F$) value, as a function of $j$.
Solid (dotted) curves are calculated with BD included (excluded).
Note that BD is effective only for $T\gtrsim T_F$, where the high-temperature system behaves as an intrinsic Dirac liquid with both electrons and holes contributing to transport.
  }
  \label{fig:taupower}
\end{figure}

\begin{figure*}
  \includegraphics[trim=0 12 0 15, clip]{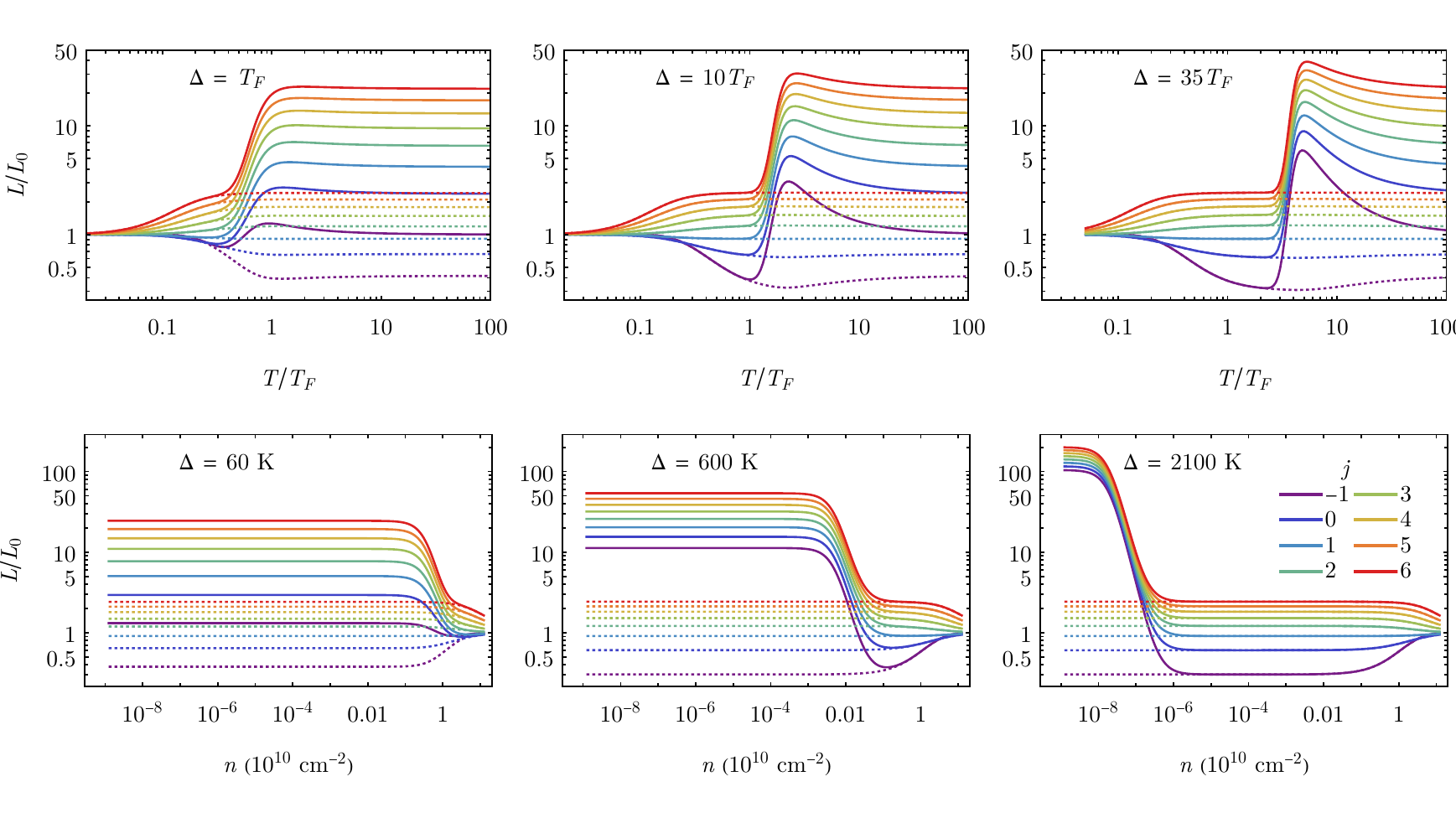}
  \caption{
    Shows the calculated $L/L_0$ for a fixed $n$ as a function of $T$ (top row) and for fixed $T=60\,\mathrm{K}$ as a function of $n$ (bottom row) in the gapped linear dispersion model with various values of the gap $\Delta$, with (solid) and without (dotted) BD. These are to be compared with the $\Delta=0$ results in Fig.~\ref{fig:taupower}.
  }
  \label{fig:taupower_gap}
\end{figure*}

Based on the theory of Sec.~\ref{sec:theory} and the motivation of understanding the reported experimental results~\cite{Crossno2016} on $L/L_0$ as a function of density and temperature, we present a detailed set of results below.
(Additional results are provided in the Appendices.)
 Our goal is to understand the large reported value of $L/L_0$ and its nonmonotonic CNP peak at a finite temperature.
 We therefore only show our calculated $L/L_0$ for various situations and parameters, both for linearly dispersing graphene systems as well as (in a few cases) for parabolically dispersing systems (imitating bilayer graphene) for the sake of comparison.
 We consider both gapless systems as well as systems with gaps.

Since we consider several different scattering mechanisms which affect the WF law quite differently, first we depict results for individual scattering processes separately, and then combine them with varying relative scattering strengths to obtain the various predicted WF law breakdown possibilities in graphene, since the real system is typically affected by several distinct scattering mechanisms
 
\subsection{Gapless systems with single scattering mechanism}

In Fig.~\ref{fig:taupower}, we show the results for our calculated $L/L_0$ in graphene (using the theory in Sec.~\ref{sec:theory} with $\Delta=0$) for different scattering strengths $\tau(\varepsilon)= \tau_0\varepsilon^j$, where $j$, the scattering exponent, characterizes the scattering mechanism.
The details of $j$ would depend on the actual microscopic scattering mechanism (e.g.\ Appendix~\ref{sec:microscopic}), and in principle, $j$ could vary even for the same physical scattering mechanism depending on the energy.
The calculated $L/L_0$ depends on $T$ and $T_F$, where the Fermi temperature $T_F$ depends on the carrier density through $T_F= (\hbar v_F/k_B) (\pi n)^{1/2} \sim 135 (n/(10^{10}\,\mathrm{cm}^{-2}))^{1/2}\;\mathrm{K}$.
The carrier density $n$ defining $T_F$ is the actual doping density or the residual puddle density defining the CNP (approximately of the order of the density regime defining the conductivity minimum plateau regime around CNP, which varies between $10^8\text{--}10^{11}\,\mathrm{cm}^{-2}$ in graphene depending on the sample quality)~\cite{Adam2007,Rossi2009,Rossi2008,Martin2008}.
 We show results with and without bipolar diffusion to emphasize the role of the Dirac point where of course the bipolar diffusion would become singular at the putative critical point (had there been no puddles at all).
We ignore all quantum critical and hydrodynamical effects arguing them to be unimportant in real graphene samples which are always dominated by puddles suppressing the Dirac point.
 Obviously, the effect of bipolar diffusion is suppressed (enhanced) at higher (lower) carrier density $n$ (with $T_F\sim n^{1/2}$), as the chemical potential moves away from the Dirac point.  Bipolar diffusion is also suppressed (enhanced) at lower (higher) temperatures for a fixed carrier density.

 The salient features of our results presented in Fig.~\ref{fig:taupower} are: (1) With BD, $L/L_0$ always peaks at the CNP for all scattering models and for all $T$ (in agreement with Ref.~\cite{Crossno2016}), but not necessarily so without BD; (2) the peak value of $L/L_0$ (nominally at the CNP) increases with increasing $j$; (3) for $j\geq 0$ with BD, $L/L_0$ increases with $T$ at higher $T$ ($>0.5 T_F$) after occasional nonmonotonicity at low $T$; (4) at high $T$ ($>T_F$), $L/L_0$ saturates to a $j$-dependent constant; (5) for $T=0$, $L/L_0$ always approaches unity, obeying the WF law precisely; (6) for $j\geq 0$ with BD, as $n$ increases (i.e.\ increasing $T_F$), $L/L_0$ decreases monotonically from its $n=0$ peak value, approaching unity at large $n$ (where $T\ll T_F$ applies).

 These features are all physically sensible in a gapless system with both electrons and holes.
At any finite $T$, the BD effect is the strongest at $n=0$, thus strongly enhancing $L/L_0$ at the CNP, and increasing $T$ enhances this effect monotonically, thus enhancing $L/L_0$.
 At $T=0$, the system must obey $L/L_0 =1$ (i.e.\ the WF law) at any finite $n$ (which is always ensured by the existence of puddles).
Therefore, as long as the limit $T=0$ is taken before the limit $n=0$, the WF law must always be obeyed in our model.
 At high $T$, $T/T_F\gtrsim 1$, the system becomes nondegenerate, leading to a modified scattering-dependent WF law, with a universal $L/L_0$, which now depends on the scattering exponent $j$.
 These values of saturated $L/L_0$ are shown in the inset of Fig.~\ref{fig:taupower}(b) as a function of $j$.
In particular, the value of $L/L_0\sim 2.4$ for $j=0$ and $\sim 4.2$ for $j=1$ has been discussed in Ref.~\cite{Yoshino2015}.

 We note that the results of Fig.~\ref{fig:taupower} are generic to all gapless systems with there being nothing special about linearly dispersing Dirac fluids.
 For example, we show the corresponding results for the gapless parabolic dispersion in Appendix~\ref{sec:parabolic}, and the results for the linear and the parabolic cases are qualitatively identical.
 This is expected because the key physics is gaplessness and the existence of both electrons and holes, with the nature of the energy dispersion itself being just a minor quantitative detail.
 We thus expect monolayer and bilayer graphene to manifest similar qualitative WF behavior.

 The same is, however, not true if the system is strictly a unipolar one-component (i.e.\ single-band) system with just electrons (or just holes).
 This is understandable since such a single-band system does not have any bipolar diffusion.
 We depict the results for a single linearly dispersing band in Appendix \ref{sec:singlelinear}, and it is clear that these results are qualitatively similar to the results without BD in Fig.~\ref{fig:taupower} (dotted curves).
The peak in $L/L_0$ at the CNP is thus a direct effect of bipolar diffusion and does not manifest when bipolar diffusion is neglected.


 Our results presented so far can explain only a part of the data of Ref.~\cite{Crossno2016}:  we cannot explain the important experimental observation of a temperature-nonmonotonic CNP peak in $L/L_0$ which maximizes at a large value ($\sim 20$) at a finite $T \sim 60\,\mathrm{K}$.
 Although $L/L_0$ can be large ($>20$) in the nondegenerate high-$T$ regime for large $j$ in the gapless BD theory (Fig.~\ref{fig:taupower}), they never manifest the nonmonotonicity in $T$.
Combining different mechanisms in a temperature-dependent way does produce some nonmonotonicity.
However, to produce the large, sharp peak at finite $T$, we need to fine-tune the theory so that the scattering is dominated by some large-$j$ mechanisms for small $T$, then suddenly changes to be dominated by small-$j$ mechanisms around $T\sim 70\text{--}90\,\mathrm{K}$, which is unlikely to be happening in the experiment.
(We mention that such a possibility of some new scattering mechanism to have a sudden onset is a possibility that cannot be ruled out no matter how unlikely it is, but we do not consider such an unlikely scenario to be a generic explanation for the observations in Ref.~\cite{Crossno2016}.)
Therefore, we conclude that the results of Ref.~\cite{Crossno2016} cannot be well-explained by the gapless model even after the full inclusion of bipolar diffusion and multiple scattering processes.


\subsection{Gapped systems with single scattering mechanism}

 We now consider, purely phenomenologically, whether the existence of an energy gap at the Dirac point for the two-band system can better explain the experiment \cite{Crossno2016}.
 We note that the symmetry protecting the gaplessness of the graphene Dirac point may be broken by the hBN substrate \cite{DeanPrivate,Hunt2013,RibeiroPalau2018,Finney2019,Yankowitz2018}.
 Thus, the emergence of a Dirac point energy gap for graphene on hBN substrates is a feasible idea although we are by no means claiming that this is what happened in Ref.~\cite{Crossno2016}.
 Our interest is to investigate how an energy gap affects the WF law and the value of $L/L_0$ in graphene.
 Our results can be tested in future experiments by deliberately inducing a Dirac point gap.

Introducing a gap $\Delta$ at the Dirac point, we recalculate $L/L_0$ for different values of $\Delta$ as a function of $T$ and $n$.
In Fig.~\ref{fig:taupower_gap} we show the results for three representative values of $\Delta$, as a comparison to the $\Delta=0$ case in Fig.~\ref{fig:taupower}.
It is clear that finite $\Delta$ tends to produce a maximum in $L/L_0$ at a finite value of $T$ (for fixed $n$ or $T_F$), which can be very large for $\Delta\gg T_F$, but the peak in $L/L_0$ as a function of density still remains at the CNP ($n\sim 0$).
This is qualitatively consistent with the experimental finding in Ref.~\cite{Crossno2016} that the peak in $L/L_0$ happens at a finite $T$ in contrast to the mostly monotonic $T$-dependence of $L/L_0$ for $\Delta=0$ (Fig.~\ref{fig:taupower}(a)).
 This peak value (now at a finite $T/T_F$) grows with increasing scattering exponent $j$ in a qualitatively similar manner as for $\Delta =0$.
 We emphasize that this nonmonotonicity in temperature occurs only in the bipolar diffusion theory (and not if only electrons or only holes are considered), showing the key role of both electrons and holes in creating this finite-temperature peak in $L/L_0$ for finite $\Delta$.
 We also emphasize that the peak is strongly pronounced for large $\Delta$ and the maximum value of $L/L_0$ can be large, consistent with Ref.~\cite{Crossno2016}, even for small $j$.
We note that very similar behavior occurs also for gapped electron-hole 2-band parabolic systems, and the corresponding results are shown in Appendix \ref{sec:parabolic}.

\subsection{Multiple scattering mechanisms}

\begin{figure*}
  \includegraphics[trim=7 5 0 0, clip]{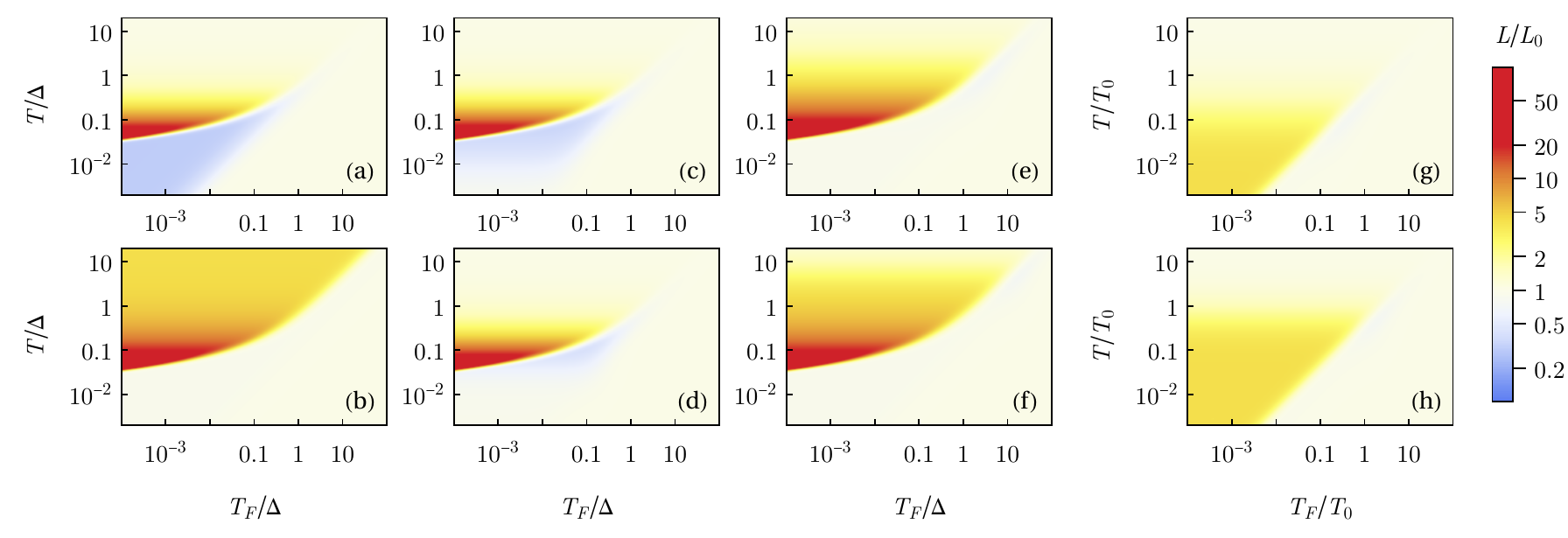}
  \caption{
    False color ``phase diagrams'' for $L/L_0$ as a function of temperature and density ($T_F\sim n^{0.5}$) with various combinations of parameters.
    (a) Gapped system with pure $j=-1$ scattering ($C=0$).
    (b) Gapped system with pure $j=1$ scattering ($A=B=0$).
    (c--f) Gapped systems with combined scattering mechanisms, with parameters $(\Delta^2A/C,\Delta^3B/C)=$ 
    $(10^3,10^3)$,
    $(10,10^3)$,
    $(0.1,10^{-3})$,
    $(10^{-3},10^{-3})$, respectively.
    (g--h) Gapless systems with combined scattering mechanisms, with parameters $(T_0)^2A/C=$ 
    $10$ and 
    $0.1$, respectively, where $T_0=(C/B)^{1/3}$ is a reference temperature scale (the scale at which phonon dominates over long-range disorder). 
    Note that the gap is the key to having the large peak of $L/L_0$ in $T$, which does not happen for the gapless case.
}
  \label{fig:phase}
\end{figure*}

Having found empirically a phenomenological resolution of the peculiar unexplained features of the Crossno WF breakdown behavior (i.e.\ large $L/L_0$ along with the peak at finite $T$), we now proceed to produce a theoretical ``phase diagram'' for $L/L_0$ in the relevant parameter space with multiple scattering mechanisms combined.
The likely generic scenario in experimental samples is the presence of multiple scattering mechanisms with varying strengths as functions of temperature, density, and energy gap.
 In doing so, however, we run into the problem of a huge over-abundance of our results with far too many theoretical parameters: $T$, $n$, $\Delta$, and the parameter $\tau_0$ for each scattering exponent $j$.
 Obviously, even if we produce such many-parameter phase diagrams, they will be useless to convey any real understanding, so we must make a choice.
 This is particularly true since all the operational scattering mechanisms (and their quantitative details) in graphene layers used in Ref.~\cite{Crossno2016} may not necessarily be known, and as is obvious from the results presented so far, the behavior of the WF breakdown associated with the dependence of $L/L_0$ on $n$ and $T$ depends on the scattering mechanism (in addition to $\Delta$, $T$, $n$) through the scattering exponent $j$.
 If several different scattering mechanisms are involved, $L/L_0$ will be a complex combination of all possible scattering mechanisms with possible temperature and density dependence of $\tau_0$ for each $j$.
 To move forward, we must make some decisions on the relevant scattering mechanisms and their relative strengths in graphene.

 From transport measurements~\cite{Zuev2009,Efetov2010,Tan2007}, it is known that the dominant transport mechanisms in graphene are short-range and long-range disorder scattering and acoustic phonon scattering.
 To keep things under control, we consider \emph{only} these three scattering mechanisms phenomenologically from now on.
 Fortunately, the nature of these scattering mechanisms is well-understood from prior work~\cite{Hwang2008a,Hwang2009}, and can be represented as:
\begin{equation}
  \tau(\varepsilon)=\frac{\tau_0}{A\varepsilon+BT\varepsilon+\frac{C}{\varepsilon}}\,.
\end{equation}
Here $A$, $B$, $C$ are parameters representing respectively the strengths of short-range disorder, acoustic phonon scattering, and long-range Coulomb disorder scattering.
 (Note that the $A$ or the $B$ term, being proportional to energy, can also be thought to represent the electron-electron scattering in intrinsic graphene, but we do not believe it to be useful to pursue this direction as electron-electron scattering is likely to be unimportant for graphene transport experiments leading to the WF law or its failure.)  The relative values of $A$, $B$, $C$ are strongly sample-dependent and unknown in general, and we are not interested in any detailed quantitative modeling anyway.
 Our goal is a minimal theory that is qualitatively consistent with the experiment of Ref.~\cite{Crossno2016}.
 Therefore, we focus on producing $L/L_0$ phase diagrams in $T$-$n$ space by varying  $\Delta$, $A$, $B$, $C$ to check whether the findings of Ref.~\cite{Crossno2016} can be explained qualitatively.
 Note that the Lorenz number of the model reduces to the $j=-1$ case if $C=0$, and the $j=1$ case if $A=B=0$.
 Also note that only the relative strengths of $A$, $B$, and $C$ affect $L/L_0$, so we will express the results as functions of the parameters $A/C$ and $B/C$ for $C\neq 0$.

 In Fig.~\ref{fig:phase} we provide a series of calculated false color ``phase diagrams'' for $L/L_0$ by varying $A$,$B$,$C$ using $T/\Delta$ and $T_F/\Delta$ as the dimensionless temperature and density, respectively ($T_F\sim n^{0.5}$).
 First, we show in Fig~\ref{fig:phase}(a)--(b) two phase diagrams for pure $j=-1$ ($C=0$, at least one of $A,B\neq 0$) and pure $j=1$ ($A=B=0$, $C\neq 0$), respectively.
 One may think of these two results as representing the situation dominated by either short-range disorder or phonon, and the situation dominated by long-range Coulomb disorder, respectively.
 Next, we show four representative combinations of $A$, $B$, $C$ in Fig.~\ref{fig:phase}(c)--(f).
 It is apparent that the presence of a gap leads to a maximum in $L/L_0$ at finite $T\sim 0.1 \Delta$ for $T_F\lesssim 0.1\Delta$, and the peak value of $L/L_0$ can be very large.
 This is consistent with Ref.~\cite{Crossno2016}.
 Although varying $A$, $B$, and $C$ changes some quantitative details (especially the size and shape of the peak), the qualitative behavior is essentially the same.
 On the other hand, we show in Fig.~\ref{fig:phase}(g)--(h) two combinations of $A$, $B$, $C$ for the $\Delta=0$ case.
 Although there is nonmonotonicity of $L/L_0$ in $T$ for some $T_F$, the peak value is always small ($L/L_0<4.2$), showing that the gap $\Delta$ is the key for having the large peak with $L/L_0>10$ in Ref.~\cite{Crossno2016}.

\begin{figure*}
  \includegraphics[trim=0 12 0 10, clip]{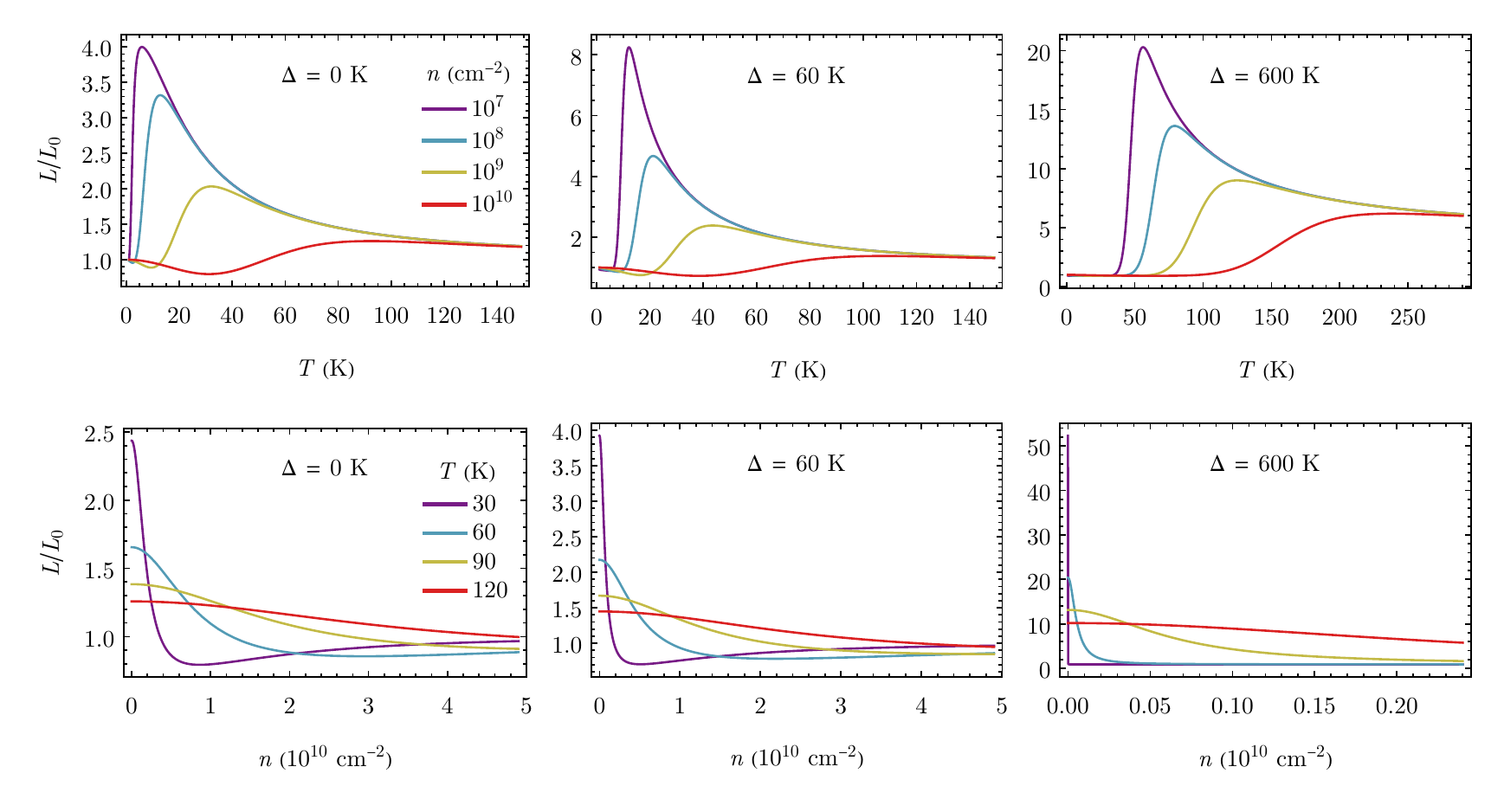}
  \caption{
    Shows the calculated $L/L_0$ for fixed $n$ as a function of $T$ (top row) and for fixed $T$ as a function of $n$ (bottom row) with various values of $\Delta$ with combined scattering mechanisms.
    The parameters are $A/C=2.78\times10^{-5}\,\mathrm{K}^{-2}$, $B/C=4.63\times10^{-6}\,\mathrm{K}^{-3}$ for the $\Delta=0\,\mathrm{K}$ and $\Delta=60\,\mathrm{K}$ figures; and $A/C=2.78\times10^{-9}\,\mathrm{K}^{-2}$, $B/C=4.63\times10^{-12}\,\mathrm{K}^{-3}$ for the $\Delta=600\,\mathrm{K}$ figures.
    Note that only the results with finite gaps show qualitative agreement with the observation of Ref.~\cite{Crossno2016}.
  }
  \label{fig:cuts}
\end{figure*}

Finally, in Fig.~\ref{fig:cuts}, we present some line plots for $L/L_0$ as functions of $T$ (for fixed $n$) and $n$ (for fixed $T$) for both gapless and gapped systems, with specific choices for $A$, $B$, $C$, showing that our gapped results indeed agree qualitatively with the results of Ref.~\cite{Crossno2016}, but not our gapless result.
Although the quantitative details depend on all the parameter details (i.e.\ $A$, $B$, $C$, $\Delta$, the definition of CNP as constrained by puddles), these details are unknown in the experiments, and it makes little sense for us to adjust parameters to obtain fine-tuned fits to Ref.~\cite{Crossno2016}, which would be misleading.
In fact, whether the experimental samples indeed have an energy gap is unknown, and there may be some other scattering mechanisms contributing to $L/L_0$ as well, such as the phonons in the substrate~\cite{Hwang2013}.
We emphasize that our key finding is that without an energy gap, no plausible explanation for the temperature-dependent nonmonotonicity in the reported $L/L_0$ exists in the theory, and the competing alternate theory of Ref.~\cite{Lucas2016} has no explanation whatsoever for the finite temperature peak in $L/L_0$~\cite{LucasPrivate}.
We note that the functional dependence of the results on $A$,$B$,$C$, $\Delta$ as well as $T$ and $n$ is quite complex, and no simple summary of the theoretical results is possible except to emphasize that for finite $\Delta$, it is possible to find regimes of $n$ and $T$ where our theory qualitatively accounts for the breakdown of the WF law reported in Ref.~\cite{Crossno2016}.
Only future experiments, perhaps using deliberately gapped graphene samples, can decisively settle the applicability of our theory to the resolution of the puzzle posed by the data of Ref.~\cite{Crossno2016}.  

We have also carried out microscopic model calculations with phonon and short-range disorder scattering just to establish that our phenomenological model using a parametrized $\tau(\varepsilon)$ is a sensible approach, which is of course expected.
These results are presented in Appendix~\ref{sec:microscopic}.

\section{Conclusion}\label{sec:conclusion}

We develop a phenomenological theory for the WF behavior of graphene layers motivated by the puzzling experimental observations reported in Ref.~\cite{Crossno2016}.
Using a phenomenological scattering model with multiple scattering mechanisms (e.g. short-range disorder, long-range disorder, acoustic phonons), we calculate $L/L_0$ as a function of density and temperature in wide parameter regimes, concluding that one simple way to understand the puzzling data of Ref.~\cite{Crossno2016} is to incorporate a Dirac point energy gap in the theory.
We speculate that such an energy gap might have been unintentionally induced by the boron nitride substrate.
We cannot find any sharp finite-temperature peak in $L/L_0$, as reported in Ref.~\cite{Crossno2016} unless we assume the existence of an energy gap.
The gapless system manifests monotonic WF behavior in our theory with $L/L_0$ at the CNP increasing with temperature, eventually saturating at some model-dependent value $L/L_0 > 1$ whereas Ref.~\cite{Crossno2016} manifests a clear temperature-dependent nonmonotonicity with $L/L_0$ at the CNP showing a peak for $T\sim60\,\mathrm{K}$.
We can also explain the large observed value of $L/L_0$ in the experiment in the presence of the energy gap and scattering by disorder and phonons.
We emphasize that we are agnostic about the existence or not of an energy gap in the samples of Ref.~\cite{Crossno2016}, but it is known that the substrate hBN can indeed open a gap of the order of $10\text{--}100\,\mathrm{meV}$ in graphene under some situations.  If such a gap exists, we find that $L/L_0$ can manifest a peak at the CNP for temperatures of the order of $10\%$ of the gap, and the peak value could be $10\text{--}40$ as observed experimentally.

We mention that the measured conductivity in Ref.~\cite{Crossno2016} is large, indicating the high quality of the graphene samples.
It is well-known that the main resistive scattering mechanisms in graphene are impurity scattering and acoustic phonon scattering although both are weak in the high-quality samples used in Ref.~\cite{Crossno2016}~\cite{DasSarma2011}. We emphasize, however, that a $T^2$ temperature dependence in the measured resistivity was not reported in Ref.~\cite{Crossno2016} or in any other graphene transport measurements, implying the absence of any observable electron-electron scattering in transport.
Phonon scattering is weak, but is generally the leading temperature-dependent scattering mechanism as it produces a linear-in-temperature increasing resistivity~\cite{DasSarma2011,Efetov2010,Hwang2008a}.
We emphasize, however, that the scale of phonon scattering here is determined by the effective Bloch-Gr\"uneisen temperature, $T_\text{BG} =2c_s k_F$, where $c_s$ and $k_F$ are the phonon velocity and the Fermi wavenumber, respectively (Appendix~\ref{sec:microscopic}), and not by the Debye temperature $T_D$, as discussed in details in Refs.~\cite{Efetov2010,Hwang2008a}.
The reason is that transport is determined mainly by $2k_F$ scatterings of electrons across the Fermi surface, which is dominated by $2k_F$ phonons with energies corresponding to $T_\text{BG}$.
It is only when $T_\text{BG}$ exceeds $T_D$, as in normal metals, is the phonon scattering cutoff determined by $T_D$.
In low-density materials such as graphene, $T_D \sim 10^3\,\mathrm{K}$ and $T_\text{BG} \sim 10\text{--}100\,\mathrm{K}$, and thus, phonon scattering can be appreciable already at $10\text{--}50\,\mathrm{K}$.
Typically, the phonon equipartition regime applies for $T >T_\text{BG}/5$, and therefore, phonon scattering is an important source of the temperature-dependent resistivity in graphene for $T >T_\text{BG}/5$, which is the main regime of interest in the experiments of Ref.~\cite{Crossno2016}.
This is consistent with the measured temperature-dependent resistivity presented in Ref.~\cite{Crossno2016}. This is why phonon scattering must be included in the theory of the WF law in graphene as we do in the current work, and the very high Debye temperature $\sim 10^3\,\mathrm{K}$ of graphene is irrelevant.

It may be useful to point out that graphene on hBN is a very weakly interacting system with the effective electron-electron interaction strength associated with Coulomb coupling being only $0.4$ (this is the so-called graphene fine-structure constant or the dimensionless Wigner-Seitz parameter~\cite{DasSarma2011}) by virtue of the large lattice dielectric constant of boron nitride. 
Thus, electron-electron interaction in extrinsic graphene is negligible~\cite{Barnes2014}.  Compared with regular normal metals (e.g.\ Al), where the dimensionless Coulomb coupling is $\sim 6$, graphene interaction effects are tiny.  
Given that electron-electron interaction effects hardly show up in the WF behavior of any normal metals, it may be a challenge to ever see interaction or hydrodynamical effects in bulk graphene. 
The only hope is to somehow access the graphene Dirac point, i.e.\ study intrinsic graphene, but this would require an extreme fine-tuning in density and temperature and disorder so that the temperature is high enough to overcome the impurity scattering and puddle effects, but low enough to avoid phonon scattering.  
Such a regime is difficult to find and would be very small in the parameter space, particularly since phonons become important at low temperatures for low carrier densities. 

Our theory by no means explains all quantitative aspects of the data in Ref.~\cite{Crossno2016}, but we do explain the most puzzling qualitative features arising from bipolar diffusion in a gapped system.  For example, Ref.~\cite{Crossno2016} seems to indicate a stronger decrease of $L/L_0$ with increasing carrier density away from the CNP than our theory gives, which may be a possible signature of hydrodynamic interaction effects which tend to suppress $L/L_0$~\cite{Ahn2022,Lucas2018}, but it is difficult to conclude anything decisively without more extensive data on many more samples~\cite{KimPrivate}.

More experimental work is necessary to settle whether a gap exists or not, and whether our theory is the qualitative explanation for the observations in Ref.~\cite{Crossno2016}~\cite{KimPrivate}.
Experiments that induce a gap deliberately and measure how the behavior of $L/L_0$ depends on the gap can help considerably in this respect and validate our theoretical predictions.
In addition, our work shows that both linear and parabolic systems behave similarly, and therefore monolayer graphene and bilayer graphene should reflect qualitatively similar Wiedemann-Franz behavior in similar parameter regimes, so our work also gives a future direction for experimental work on verifying such WF law behavior for bilayer graphene.

We emphasize that our theory makes clear predictions for future experimental work:
Open a gap in graphene by aligning the hBN substrate appropriately, and the effective $L/L_0$ will be enhanced as predicted in our theory. Our predicted WF behavior in graphene should be tested in future experiments.

There are many open questions that should be studied in future works.
We are, however, unsure that these questions should be explored before the experimental situation clarifies~\cite{KimPrivate}.
The most important shortcoming of our theory is the neglect of electron-electron interactions arguing that the system is unlikely to be in the hydrodynamical regime since the electron-impurity and/or electron-phonon scatterings likely dominate over the electron-electron scattering in experimental graphene samples with the Dirac point criticality inaccessible experimentally. 
We also hypothesize, based on our microscopic calculations of the electron-electron scattering, that Ref.~\cite{Crossno2016} is most likely not in the hydrodynamical regime.
This, however, requires further study, perhaps by generalizing the detailed hydrodynamical theory of Ref.~\cite{Lucas2018} to the electron-hole 2-component situation.
The main problem is that electron-electron scattering by itself always suppresses the $L/L_0$ ratio in metallic Fermi liquids~\cite{Lucas2018,Ahn2022}, and therefore, it is unclear how the puzzling enhancement of $L/L_0$ in Ref.~\cite{Crossno2016} can be explained by interactions.
In addition, electron-electron interactions would increase monotonically as a function of temperature~\cite{Li2013} and it is a challenge to find a temperature-nonmonotonicity caused by electron-electron interactions, as indeed Ref.~\cite{Lucas2016}, using a phenomenological hydrodynamical theory with adjustable parameters, only found increasing $L/L_0$ with increasing temperature.
We believe that the main physics observed in Ref.~\cite{Crossno2016} is the physics of bipolar diffusion, and if the system is gapless, the maximum $L/L_0$ should be $<5$. 

\section*{Acknowledgment}

The authors thank Professors Philip Kim and Andrew Lucas for useful discussions.  This work is supported by the Laboratory for Physical Sciences.

\appendix

\section{Parabolic dispersion}\label{sec:parabolic}

Here we modify Eq.~(\ref{eq:dispersion}) to parabolic dispersion
\begin{align}
  \varepsilon_{+}(\mathbf{p})&=+ \frac{|\mathbf{p}|^2}{2m}\nonumber\\
  \varepsilon_{-}(\mathbf{p})&=- \frac{|\mathbf{p}|^2}{2m}-\Delta\,,
\end{align}
where $\Delta$ is the size of the gap and $m$ is the effective mass.
 For better comparison with the linear dispersion case in the main text, we consider the same spin and valley degeneracy as in graphene, so that the density of state is $D_\pm(\varepsilon)=2m/(\pi\hbar^2)$, and we choose $m=60\,\mathrm{K}k_B/v_F^2$, where $v_F$ is the Fermi velocity of graphene.

\begin{figure}
  \includegraphics[trim=5 15 0 0, clip]{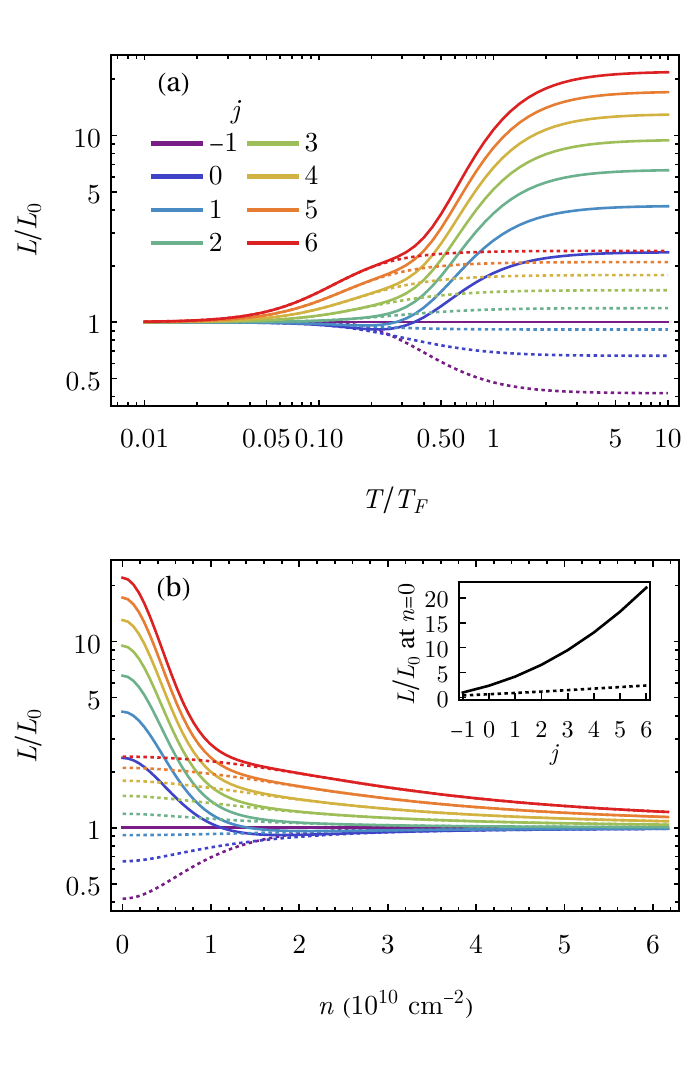}
  \caption{
    Shows the calculated $L/L_0$ for (a) fixed $n$ as a function of $T$ and for (b) fixed $T=60\,\mathrm{K}$ as a function of $n$ in the gapless parabolic dispersion model for a scattering rate $\tau(\varepsilon)=\tau_0\varepsilon^j$, with (solid) and without (dotted) BD.
    These are to be compared with the linear dispersion result in Fig.~\ref{fig:taupower}. The qualitative similarity shows that the exact band structure is not important.
}
  \label{fig:gaplessparabolic}
\end{figure}

\begin{figure*}
  \includegraphics[trim=0 12 0 15, clip]{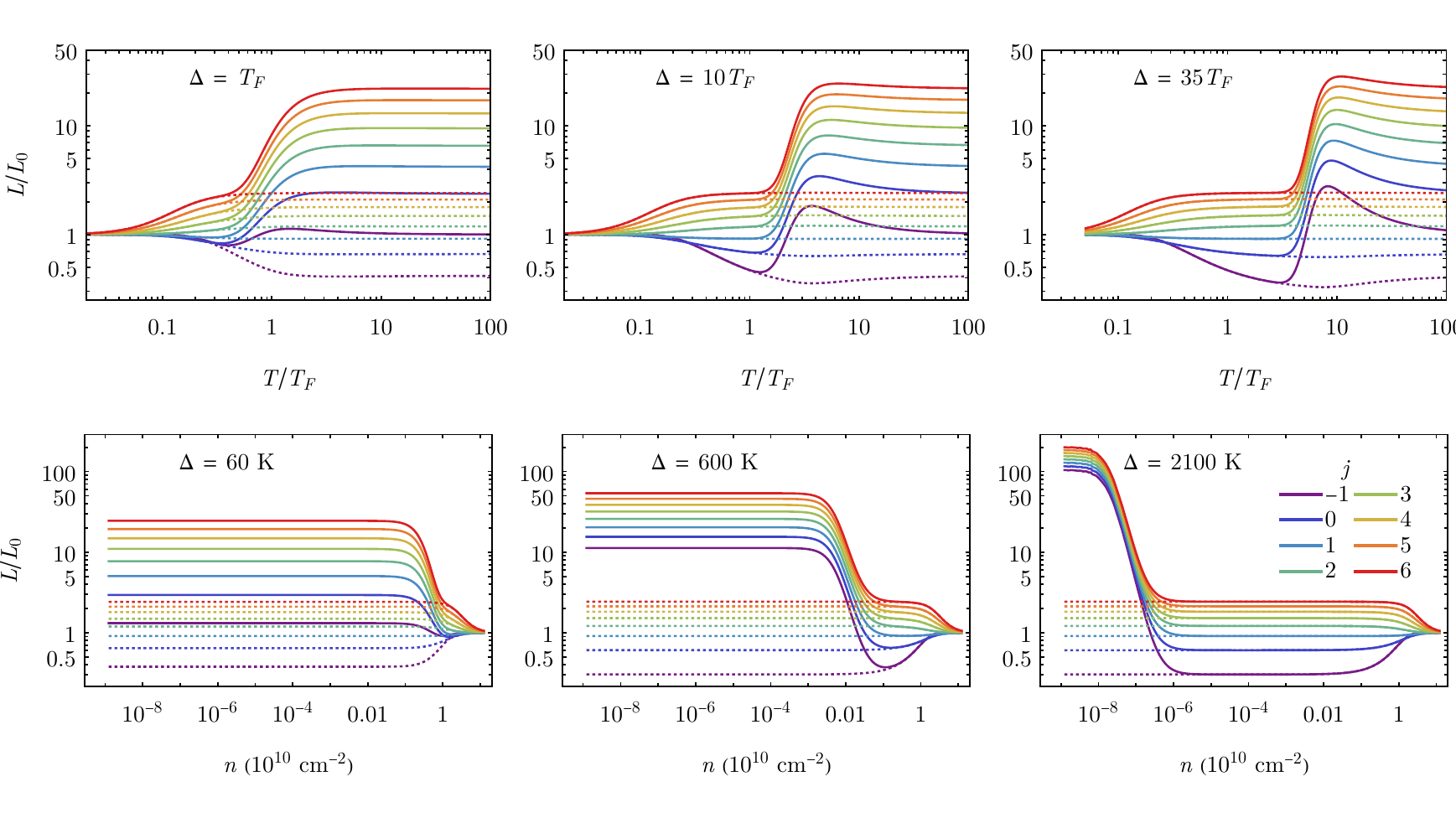}
  \caption{
    Shows the calculated $L/L_0$ for a fixed $n$ as a function of $T$ (top row) and for fixed $T=60\,\mathrm{K}$ as a function of $n$ (bottom row) in the gapped parabolic dispersion model with various values of the gap $\Delta$, with (solid) and without (dotted) BD.
    These are to be compared with the linear dispersion result in Fig.~\ref{fig:taupower_gap}. The qualitative similarity shows that the exact band structure is not important.
  }
  \label{fig:parabolic_gap}
\end{figure*}

 Figs.~\ref{fig:gaplessparabolic} and \ref{fig:parabolic_gap} show the calculated $L/L_0$ with $\Delta=0$ and $\Delta>0$, respectively, for $\tau(\varepsilon)=\tau_0\varepsilon^j$. Note the qualitative similarity with the corresponding linear dispersion results: Figs.~\ref{fig:taupower} and \ref{fig:taupower_gap} in the main text. This shows that only the gaplessness, rather than the exact dispersion near the Dirac point, affects the qualitative results.
 This validates our simple choice of the band structure near the gap opening, intended to model the unknown band structure induced by the hBN substrate.
 In addition, since the low-energy band structure of bilayer graphene can be modeled by gapless parabolic bands~\cite{DasSarma2011}, we expect that similar behavior of $L/L_0$ discussed in the main text also exists in bilayer graphene.

\section{Single linear band}\label{sec:singlelinear}

To address the key importance of bipolar diffusion by both electrons and holes, we arbitrarily consider the hypothetical single-band case.
Here we set $\varepsilon_{+}(\mathbf{p})=+ v_F |\mathbf{p}|$ without the $\varepsilon_{-}$ band in Eq.~(\ref{eq:dispersion}).
This can be considered as just calculating $L_+=\kappa_+/(\sigma_+T)$ in Sec.~\ref{sec:theory} with the second term in Eq.~(\ref{eq:consistent}) dropped, or can be equivalently considered as taking $\Delta\to\infty$ in the main text.
Note that as $\Delta$ becomes larger, the effective region of BD shrinks towards smaller $n$.
And finally, as we take the limit $\Delta\to\infty$, the BD becomes effective only at $n=0$, which is meaningless since there is no carrier in the system at that point.
Therefore, for either approach, there is no BD here.

\begin{figure}
  \includegraphics[trim=5 15 0 15, clip]{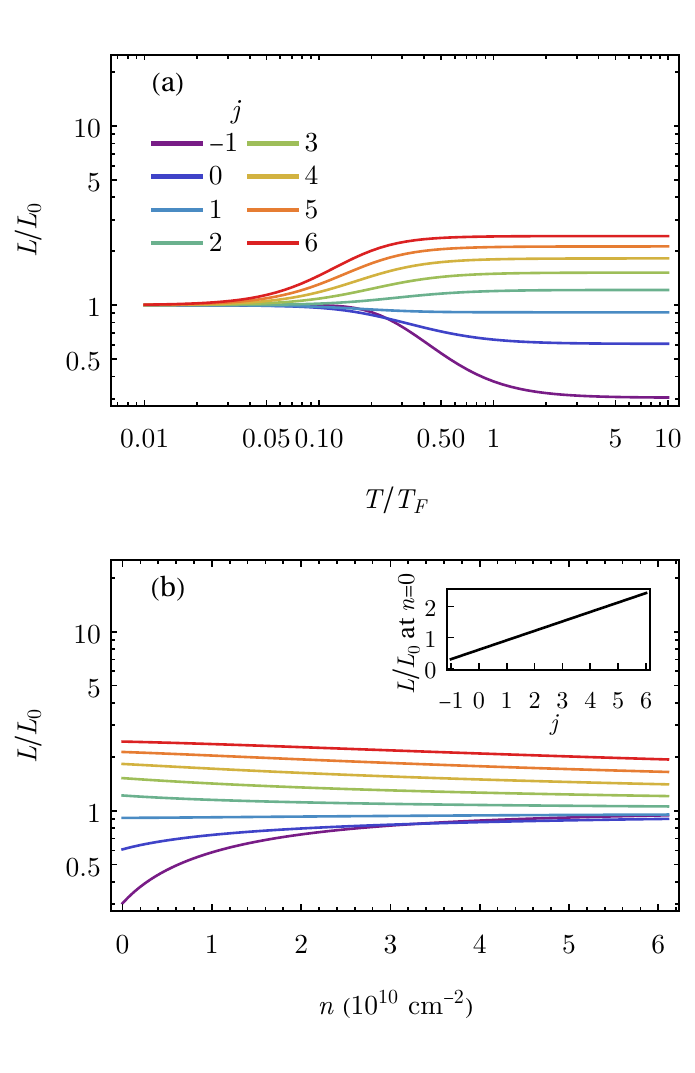}
  \caption{
    Shows the calculated $L/L_0$ for (a) fixed $n$ as a function of $T$ and for (b) fixed $T=60\,\mathrm{K}$ as a function of $n$ in the single linear band model for a scattering rate $\tau(\varepsilon)=\tau_0\varepsilon^j$.
    Note that there is no large enhancement in $L/L_0$, and the result is qualitatively similar to the no-BD result (dotted curves) of Figs.~\ref{fig:taupower} and \ref{fig:taupower_gap}.
  }
  \label{fig:singlelinear}
\end{figure}

Fig.~\ref{fig:singlelinear} shows the calculated $L/L_0$ for $\tau(\varepsilon)=\tau_0\varepsilon^j$. Note the lack of large enhancement of $L/L_0$, and the qualitative similarity with the no-BD result (dotted curves) of Figs.~\ref{fig:taupower} and \ref{fig:taupower_gap} in the main text.
(The different behavior near $n=0$ can be explained by the difference in the consistency condition determining the chemical potential.)
This shows that having both the conduction and the valance band, and hence the BD, is essential in producing the large enhancement of $L/L_0$ described in the main text.
This enhancement is what distinguishes graphene from regular metals.

\section{Microscopic calculations}\label{sec:microscopic}

In this appendix, we calculate the electrical ($\tau_\sigma$) and thermal ($\tau_\kappa$) relaxation times for the band structure of Eq.~(\ref{eq:dispersion}) directly from the Boltzmann equation formalism with the relaxation time approximation, without using the phenomenological model for the relaxation time as in the main text, and obtain results of $L/L_0$ from these calculated relaxation times.
Since we split the system into the conduction band and the valance band, we only consider the conduction band for the equations below. The equations for the valance band are obtained by particle-hole symmetry.

We follow the approach of Ref.~\cite{Lavasani2019}, starting from the Boltzmann equation
\begin{equation}
  \frac{\partial f}{\partial t} + \dot{\mathbf{r}}\cdot \frac{\partial f}{\partial \mathbf{r}} + \dot{\mathbf{k}}\cdot \frac{\partial f}{\partial \mathbf{k}} = \mathcal{I}\{f\}\,,
\end{equation}
where $f(\mathbf{r},\mathbf{k},t)$ is the distribution of electron wave packets at position $\mathbf{r}$, momentum $\mathbf{k}$ (we set $\hbar=k_B=1$ for brevity), and time $t$;
The semiclassical equations of motion are
\begin{align}
  \dot{\mathbf{r}}&=v_F \mathbf{\hat k}-\pi e \mathbf{E}\times\mathbf{\hat z}\;\delta^2(\mathbf{k})\label{eq:rdot}\\
  \dot{\mathbf{k}}&=-e\mathbf{E}\,,
\end{align}
where $\mathbf{E}$ is the applied electric field.
The second term of Eq.~(\ref{eq:rdot}) comes from the Berry curvature at the Dirac point~\cite{Xiao2010}, which we will neglect because it has no effect on the longitudinal conductivity in the linear response regime.
The collision integral $\mathcal{I}$ is given by
\begin{align}
  \mathcal{I}\{f\}(\mathbf{k})=-\int \frac{d^2\mathbf{k}'}{(2\pi)^2} \big[&S(\mathbf{k},\mathbf{k}')f(\mathbf{k})(1-f(\mathbf{k}'))\nonumber\\
                                                                          +&S(\mathbf{k}',\mathbf{k})f(\mathbf{k}')(1-f(\mathbf{k}))\big]\,,
\end{align}
where $S(\mathbf{k},\mathbf{k}')$ is the differential scattering rate from the state with momentum $\mathbf{k}$ to $\mathbf{k}'$.

By assuming $f=f_0+\delta f$ for a small perturbation $\delta f$ near the local equilibrium $f_0$, the Boltzmann equation can be linearized as
\begin{equation}
  \frac{\partial\delta f}{\partial t} - \dot{\mathbf{r}}\cdot\left(e\mathcal{E} + \frac{\varepsilon-\mu}{T} \nabla T\right)\frac{\partial f_0}{\partial\varepsilon}=\mathcal{I}\{f_0+\delta f\}\,.
\end{equation}
Using the relaxation time ansatz for steady-state solutions of $\delta f$ (as in \cite{Lavasani2019}):
\begin{equation}
  \delta f=\dot{\mathbf{r}}\cdot\left(\tau_\sigma e\mathcal{E} + \tau_\kappa\frac{\varepsilon-\mu}{T} \nabla T\right)\frac{\partial f_0}{\partial\varepsilon},
\end{equation}
where $\mathcal{E}=\mathbf{E}+\nabla \mu/e$, we arrive at the following equations for the relaxation times
\begin{align}\label{eq:relaxation}
  1&=\int\frac{d^2 \mathbf{k}'}{(2\pi)^2}S(\mathbf{k},\mathbf{k}')\frac{1-f_0(\varepsilon')}{1-f_0(\varepsilon)}
  \left(\tau_\sigma(\varepsilon)-\tau_\sigma(\varepsilon')\cos\theta\right)\nonumber\\
  1&=\int\frac{d^2 \mathbf{k}'}{(2\pi)^2}S(\mathbf{k},\mathbf{k}')\frac{1-f_0(\varepsilon')}{1-f_0(\varepsilon)}
  \left(\tau_\kappa(\varepsilon)-\tau_\kappa(\varepsilon')\frac{\varepsilon'-\mu}{\varepsilon-\mu}\cos\theta\right).
\end{align}
Here, $S(\mathbf{k},\mathbf{k}')$ is assumed to depend only on $k$, $k'$ and $\theta$, the angle between $\mathbf{k}$ and $\mathbf{k}'$.

In general, Eq.~(\ref{eq:relaxation}) cannot be solved exactly.
As a reasonable approximation, the relaxation time approximation considered in \cite{Lavasani2019} is used, which assumes that $\tau_{\sigma,\kappa}(\varepsilon)\approx\tau_{\sigma,\kappa}(\varepsilon')$. With such an approximation, we have
\begin{align}\label{eq:RTA}
  \frac{1}{\tau_\sigma(\varepsilon)}&=\int\frac{d^2 \mathbf{k}'}{(2\pi)^2}S(\mathbf{k},\mathbf{k}')\frac{1-f_0(\varepsilon')}{1-f_0(\varepsilon)}
  \left(1-\cos\theta\right)\nonumber\\
  \frac{1}{\tau_\kappa(\varepsilon)}&=\int\frac{d^2 \mathbf{k}'}{(2\pi)^2}S(\mathbf{k},\mathbf{k}')\frac{1-f_0(\varepsilon')}{1-f_0(\varepsilon)}
  \left(1-\frac{\varepsilon'-\mu}{\varepsilon-\mu}\cos\theta\right)\,,
\end{align}

For simplicity, we only consider short-range impurity and phonon scattering here in this microscopic calculation.
Once the scattering rates $S(\mathbf{k},\mathbf{k}')$ are obtained from the appropriate underlying Hamiltonians, it is straightforward to calculate the relaxation times.
Note that a similar microscopic calculation for the electrical conductivity has been done and discussed in detail in Ref.~\cite{DasSarma2013}.

\subsection{Short-range impurity}

We assume impurity scattering by short-range delta function potential
\begin{equation}
  V_\text{imp}(\mathbf{r})=u_0\,\delta(\mathbf{r})\,,
\end{equation}
where $u_0$ is the strength of the potential.
In addition, we assume that the impurities are distributed randomly with number density $n_\text{imp}$.
Then Fermi's golden rule leads to the scattering rate:
\begin{equation}
  S(\mathbf{k},\mathbf{k}')=2\pi n_\text{imp}\,u_0^2\,\delta(\varepsilon-\varepsilon')\cos^2\frac{\theta}{2}\,.
\end{equation}
Plugging into Eq.~(\ref{eq:relaxation}), the relaxation time can be solved exactly:
\begin{equation}
  \frac{1}{\tau_\sigma(\varepsilon)}=\frac{1}{\tau_\kappa(\varepsilon)}=\frac{n_\text{imp}u_0^2}{4v_F^2}\varepsilon\,,
\end{equation}
which has the form of the $A$ term in the phenomenological model as considered in the main text.

\subsection{Phonon}

We consider the deformation potential model for the interaction of electrons and acoustic phonons
\begin{equation}
  H_\text{e-ph}=\frac{1}{\sqrt{V}}\sum_{\mathbf{k},\mathbf{k}',\mathbf{q}}\sqrt{\frac{D^2}{2\rho\omega_q}}q\left(a_{\mathbf{q}}+a_{-\mathbf{q}}^\dagger\right)c_{\mathbf{k}}^\dagger c_{\mathbf{k}'}\delta_{\mathbf{k}-\mathbf{k}'-\mathbf{q},0}\,,
\end{equation}
where $V$ is the area of the system, $D$ is the deformation potential strength, $\rho$ is the ion density,
$\omega_q=c_s q$ is the phonon energy,
$a^\dagger$ and $c^\dagger$ are the creation operators for phonons and electrons, respectively.
The phonons are assumed to be always in thermal equilibrium, with occupation number $N_q=1/(\exp(\omega_q/T)-1)$.
We ignore the cutoff from the Debye temperature, since it is much higher than the temperature of interest, and we ignore phonon drag.
In addition, since the speed of sound $c_s\sim 20\,\mathrm{km/s}$ in graphene~\cite{DasSarma2011,Hwang2008a} is much smaller than $v_F$, we only do the calculations to the leading order in $c_s/v_F$, i.e,\ we assume the Migdal theorem, which is valid in graphene~\cite{Roy2014}. 

By Fermi's golden rule, the scattering rate is obtained
\begin{align}
  S(\mathbf{k},\mathbf{k}')=\frac{\pi D^2 q^2}{\rho\,\omega_q}\big[ &N_q \delta(\varepsilon-\varepsilon'+\omega_q)\nonumber\\
  +&(N_q+1)\delta(\varepsilon-\varepsilon'-\omega_q)\big]\cos^2\frac{\theta}{2}\,,
\end{align}
where $\mathbf{q}=\mathbf{k}-\mathbf{k}'$.

In this case, Eq.~(\ref{eq:relaxation}) cannot be solved exactly and we must apply the relaxation time approximation, i.e.\ using Eq.~(\ref{eq:RTA}). After some algebra, the result is
\begin{widetext}
\begin{align}
  \frac{1}{\tau_\sigma(\varepsilon)}&=\frac{2D^2}{\pi\rho v_F c_s}k^2 \frac{1}{z_0^3}\int_{-z_0}^{z_0}\frac{dz}{\sqrt{z_0^2-z^2}}\frac{e^{\eta}+1}{e^{\eta+z}+1}\frac{z}{1-e^{-z}}\left(1-\left(\frac{z}{z_0}\right)^2\right)(2z^2)\\
  \frac{1}{\tau_\kappa(\varepsilon)}&=\frac{2D^2}{\pi\rho v_F c_s}k^2 \frac{1}{z_0^3}\int_{-z_0}^{z_0}\frac{dz}{\sqrt{z_0^2-z^2}}\frac{e^{\eta}+1}{e^{\eta+z}+1}\frac{z}{1-e^{-z}}\left(1-\left(\frac{z}{z_0}\right)^2\right)\left(z_0^2-\frac{\eta+z}{\eta}(z_0^2-2z^2)\right)\,,
\end{align}
\end{widetext}
where $z_0=\frac{2k c_s}{T}$, $z=z_0\sin\frac{\theta}{2}$, and $\eta=\frac{\varepsilon-\mu}{T}$.

Note that in the phonon equipartition regime ($T\gg 2 c_s k_F=2\frac{c_s}{v_F} T_F$), only the leading order of $z_0$ and $z$ is important, so both integrals reduce to
\begin{equation}
  \int_{-z_0}^{z_0}\frac{dz}{\sqrt{z_0^2-z^2}}\left(1-\left(\frac{z}{z_0}\right)^2\right)(2z^2)\sim z_0^2\,.
\end{equation}
Therefore, in this regime, we have $1/\tau_\sigma=1/\tau_\kappa\sim k^2/z_0\sim\varepsilon T$, which has the form of the $B$ term in the phenomenological model as considered in the main text.

\subsection{Results}

\begin{figure}
  \includegraphics[trim=5 0 0 0, clip]{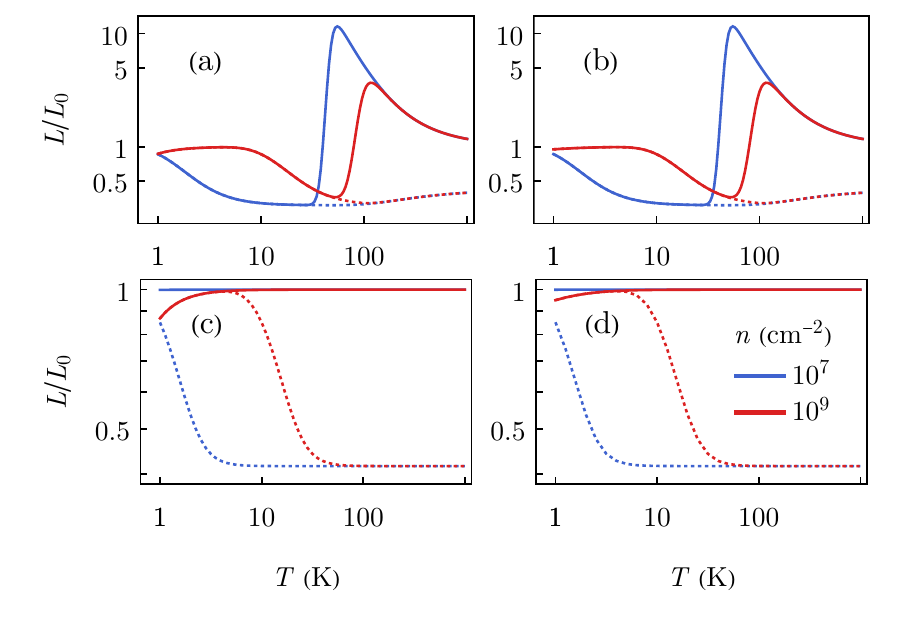}
  \caption{
    Shows the calculated $L/L_0$ for fixed $n$ as a function of $T$ for the microscopic models of linear dispersion systems, with (solid) and without (dotted) BD.
    (a) Gapped model with phonon scattering only.
    (b) Gapped model with phonon and short-range impurity.
    (c) Gapless model with phonon scattering only.
    (d) Gapless model with phonon and short-range impurity.
    We choose $\Delta=600\,\mathrm{K}$ for the gapped models and
    $n_\text{imp}u_0^2=4000\,\mathrm{K} D^2/\rho$ for the impurity model. 
    The result is consistent with the phenomenological model in the main text.
}
  \label{fig:microscopic}
\end{figure}

With the relaxation times obtained above, we use the same procedure of Sec.~\ref{sec:theory} to calculate $L/L_0$. The results are shown in Fig.~\ref{fig:microscopic}, which is consistent with the phenomenological model considered in the main text.
The only qualitative difference from the $j=-1$ model is the suppression of $L/L_0$ at low temperature when the impurity scattering is not too strong, where the scattering is dominated by phonons in the Bloch-Gr\"uneisen regime (the full details of $L/L_0$ in this regime have been discussed in \cite{Lavasani2019} for a single parabolic band, and the situation for extrinsic graphene is essentially the same as the holes and the linearity of the band play no roles at low temperature).
Since the large enhancement of $L/L_0$ in the gapped system occurs at a much higher temperature, this has no effect on the results in the main text.
Again, it is obvious that the gap is the key to the large nonmonotonic peak of $L/L_0$ at finite temperature.

\bibliographystyle{apsrev4-2}
\bibliography{references}

\end{document}